\documentclass[floatfix,superscriptaddress,twocolumn,showpacs,prb,amsmath,amssymb]{revtex4-1}
\usepackage{graphicx}
\usepackage{amssymb}
\usepackage{ulem}
\usepackage{amsbsy}
\usepackage{amsfonts} 
\usepackage{amssymb} 
\usepackage{natbib}
\usepackage{bm}

\usepackage{color}

\def\tc{$T_{\mathrm{c}}$\ }
\def\tsdw{$T_{\mathrm{SDW}}$\ }

\def\t6as{$\mathrm{(TMTSF)_{2}AsF_{6}}$\ }
\def\tmx{(TMTSF)$_{2}$(ClO$_{4}$)$_{(1-x)}$(ReO$_{4}$)$_x$\ }
\def\tmasf6sbf6{(TMTSF)$_{2}$(AsF$_{6}$)$_{(1-x)}$(SbF$_{6}$)$_x$\ }
\def\tmxns{(TMTSF)$_{2}$(ClO$_{4}$)$_{(1-x)}$(ReO$_{4}$)$_x$\ }
\def\tmc{$\mathrm{(TMTSF)_{2}ClO_{4}}$\ }

\def\tms{$\mathrm{(TMTSF)_{2}AsF_{6(1-x)}SbF_{6x}}$\ }

\def\tmttfsbf6{$\mathrm{(TMTTF)_{2}SbF_{6}}$\ }

\def\tmttfasf6{$\mathrm{(TMTTF)_{2}AsF_{6}}$\ }
\def\tmtsfasf6{$\mathrm{(TMTSF)_{2}AsF_{6}}$\ }
\def\tmttfbf4{$\mathrm{(TMTTF)_{2}BF_{4}}$\ }

\def\tmtsfreo4{$\mathrm{(TMTSF)_{2}ReO_{4}}$\ }
\def\tmno3{$\mathrm{(TMTSF)_{2}NO_{3}}$\ }
\def\tm2x{$\mathrm{(TM)_{2}X}$\ }
\def\tm2xns{$\mathrm{(TM)_{2}X}$}

\def\R{$\mathrm{ReO_{4}^{-}}$\ }
\def\C{$\mathrm{ClO_{4}^{-}}$\ }

\def\hc2{$H_{\mathrm{c2}}$\ }

\def\tmp6{$\mathrm{(TMTSF)_{2}PF_{6}}$\ }

\def\tms2x{$\mathrm{(TMTSF)_{2}}X$}
\def\tm2x{$\mathrm{(TM)_{2}}X$\ }

\def\sb{$\mathrm{SbF_{6}}$}
\def\pf{$\mathrm{PF_{6}^-}$}

\def\cl{$\mathrm{ClO_{4}^-}$}

\def\4fb{$\mathrm{BF_{4}}$}

\def\reo4{$\mathrm{ReO_{4}}$}
\def\bedtttfreo4{$\mathrm{(BEDT-TTF)_{2}ReO_{4}}$\ }
\def\et2i3{$\mathrm{(ET)_{2}I_{3}}$\,}
\def\et2x{$\mathrm{(ET)_{2}X}$\,}
\def\ket2x{$\mathrm{\kappa-(ET)_{2}X}$\,}

\def\ket2x{$\mathrm{\kappa-(ET)_{2}X}$\,}

\def\betsfecl4{$\mathrm{(BETS)_{2}FeCl_{4}}$\,}

\def\et{$\mathrm{ET}$\,}

\def\TAO{$T_{\mathrm{AO}}$\ }

\newcommand{\cstar}{c^{\star}}

\newcommand{\rhoc}{\rho_{c^\star}}

\newcommand{\rhoMin}{\rho_{\mathrm{Min}}}
\newcommand{\rhoMax}{\rho_{\mathrm{Max}}}

\newcommand{\Ohmcm}{\Omega\text{\textperiodcentered cm}}

\definecolor{darkgreen}{rgb}{0,0.6,0}

\begin{document}
\title{Influence of carrier lifetime on  quantum criticality and superconducting \tc   of \tmc }
\author{Abdelouahab   Sedeki} \affiliation{Universit\'e Dr Tahar Moulay Saida, BP 138 cit\'e ENNASR 20000, Saida, Algeria.}
\author{Pascale Auban-Senzier} \affiliation{Laboratoire de Physique des Solides (UMR 8502), Univ.Paris-Sud, 91405 Orsay, France}

\author{Shingo~Yonezawa}\affiliation{Department of Physics, Graduate School of Science,  Kyoto University, Kyoto 606-8502, Japan}

\author{Claude Bourbonnais} \affiliation{ Regroupement Qu\'eb\'ecois sur les Mat\'eriaux de Pointe and Institut Quantique,  D\'epartement de physique, Universit\'e de Sherbrooke,
Sherbrooke, Qu\'ebec, Canada, J1K-2R1}

\author{Denis~Jerome} \affiliation{Laboratoire de Physique des Solides (UMR 8502), Univ.Paris-Sud, 91405 Orsay, France}


\date{\today}

\begin{abstract}
This work presents and analyzes electrical resistivity data  on the organic superconductor (TMTSF)$_2$ClO$_4$ and their anion substituted alloys (TMTSF)$_2$(ClO$_4$)$_{1-x}$(ReO$_4$)$_x$ along the least conducting $c^\star$ axis. Nonmagnetic disorder introduced by finite size domains of anion ordering  on non Fermi liquid character of resistivity is investigated near the conditions of quantum criticality.  The evolution of the $T$-linear resistivity  term  with anion disorder shows a limited decrease in contrast  with the complete suppression of the critical   temperature $T_c$ as expected for unconventional superconductivity beyond a threshold value of $x$. The resulting breakdown of scaling between both quantities is compared to the theoretical predictions of a  linearized Boltzmann equation combined to the  scaling theory  of umklapp scattering in the presence of disorder induced pair-breaking for the carriers. \end{abstract}
\maketitle

\section{Introduction}

Quantum criticality that  becomes unstable against the emergence of superconductivity is  a  common feature of most recent unconventional superconductors. It happens when the ordering temperature of an antiferromagnetic phase (AFM), which can be metal-like, semi-metallic or insulating  ends towards zero temperature,  as a control parameter doping or pressure is varied\cite{Norman11}. Such is the situation encountered in the vast family of   pnictide superconductors\cite{Paglione10,Johnston10}. That may also be the situation  met in hole doped cuprates although there, the quantum critical point (QCP) corresponding to the termination at zero temperature of the pseudogap phase is beneath the superconducting dome of their phase diagram\cite{Taillefer10}. 
This is also found in electron doped cuprates and heavy fermions superconductors,  which both exhibit an AFM phase terminating toward  zero temperature when the electron doping level  or pressure are varied respectively in the former \cite{Dagan04} or  latter compounds\cite{Jaccard92,Mathur98}. The proximity of an AFM  Mott  insulating  phase to superconductivity is also present in layered   organic superconductors bearing much resemblance with the  phase diagram of cuprates\cite{Kanoda97,Lefebvre00,Mackenzie97}.

Most experimental investigations performed near QCP conditions have revealed that several physical properties deviate significantly from  the canonical Fermi liquid behaviour, in particular the resistivity which instead of the expected $T^2$ dependence exhibits a linear-$T$ dependence\cite{Jin11}, a puzzling feature shared by essentially all the aforementioned  unconventional superconductors close to a magnetic QCP.

Moreover, there exists another class of materials displaying both antiferromagnetism and superconductivity that has not been as much highlighted namely, the quasi-one dimensional (Q1D) Bechgaard organic superconductors\cite{Bechgaard80,Bourbonnais08}. Although superconductivity in heavy fermions\cite{Steglich79} has been discovered before that in  Q1D organics, it is in these latter materials that  superconductivity  first revealed following  the suppression  of an insulating phase under pressure which was  subsequently identified as   a spin density wave (SDW) state  \cite{Jerome80,Scott80,Andrieux81}.  A strong argument in favour of   Q1D superconductors is the relative simplicity of their electronic structure, comprising a single nearly flat Fermi surface as well as a temperature-pressure phase diagram which can be considered as  a textbook example of magnetic quantum criticality that becomes unstable to the formation of superconductivity\cite{Jerome91,Sedeki12}.

An unusual behavior of  metallic transport in \tmp6 under pressure at low temperature   has been observed in the first studies\cite{Jerome80} related to  a pronounced  $T$-linear tendency between 0.1~K and 10~K and even a downward curvature of the resistivity   in the pressure range corresponding to the vicinity of the QCP\cite{Schulz81}.
While  superconducting fluctuations were first considered as a possible interpretation in the 3D temperature regime very close to \tc where the transverse coherence length is larger than the interstack distance, paraconductivity from superconducting origin failed to account for the correct temperature dependence observed up to 10K or so\cite{Schulz81}.
Thirty years have been required  to clarify the problem  when similar behaviors of transport  have been detected in other superconductors close to a QCP in particular in cuprates and pnictides\cite{Taillefer10,Jin11}.

As far as organic superconductors are concerned,  detailed investigations of transport  at low temperature  has  been conducted  first on \tmp6 under pressure\cite{Doiron09}. In this case  the temperature dependence of the  inelastic scattering in conditions very close to the QCP  is a clear cut behavior.
$\rho(T)$ follows a $T$-linear law over a decade in temperatures above $T_c$, already visible on log-log plot of the resistivity.  The behavior is becoming quadratic as  expected in a Fermi liquid above 10~K\cite{Doiron09}.
However, as the location of the compound in the $T-P$ phase diagram  is moved away  from the QCP conditions the resistivity acquires a quadratic component even at low temperature  and becomes  fully quadratic when \tc is suppressed. The log-log plot of the resistivity reveals  a power law  ${\Delta \rho\propto T^\beta}$ with  an exponent  ${\beta}$ evolving from 1 to 2 between 11.8 and 20.8 kbar\cite{Doiron09}. 
Subsequently, an other procedure has been followed to analyse the inelastic scattering of \tmp6 namely, a  sliding fitting procedure\cite{Doiron10}  with a second order polynomial form such as $\rho(T) = \rho_0+A(T )T+B(T )T^2$.
Here, the fitting procedure enables the detection of a possible temperature dependence of the  $A$ and $B$ prefactors, whereas the value of $\rho_0$ depends on pressure only\cite{Doiron10}. This procedure has shown that the $T$-linear contribution is dominant below 10~K compared to the Fermi liquid contribution close to the QCP,   but  otherwise becomes of the order of the quadratic contribution. 

The same log-log analysis failed to provide a clear cut picture in \tmc, but the polynomial fitting procedure conducted on   the resistivity along $c^\star$, led to conclusions that  $A$ and $B$  are only weakly temperature dependent in this compound  between \tc and 15~K (though still pressure dependent)\cite{Auban11a}. 
The detailed analysis of the temperature dependent  resistivity over an extended range of pressures   has shown that $\rhoc (T)$ can thus be fitted to the  polynomial expression  $\rho_{c^\star0} +AT+ BT^2$ where all parameters are evolving under pressure but the experimental feature, $A \rightarrow$ 0 as \tc $\rightarrow 0$ is  preserved\cite{Auban11a}. 

Although a precise functional relation between $A$  and \tc is  uncertain, the salient result of these pressure studies  is the existence of a $T$-linear contribution to the resistivity     being finite only when \tc is finite, suggesting in turn the existence of a common origin for pairing and the   $T$-linear inelastic scattering\cite{Doiron10}.

 While there is no general consensus about the microscopic origin of $T$-linear resistivity appearing near a QCP\, \cite{Coleman05}, theoretical efforts have been displayed to shed some light on its origin in the context of the Q1D Bechgaard salts\cite{Sedeki12,Meier13,Shahbazi15} The   electron-electron   scattering rate, as extracted from the imaginary part of the  one-particle self-energy, has been derived by a renormalisation group procedure. The calculation that takes into account the quantum interference between  electron-electron and electron-hole scattering channels was able to link the strength of the linear resistivity term to the one of pairing as a function of pressure\cite{Sedeki12}. Another strategy has been to connect the solution of the transport Boltzmann equation to the renormalization group approach to non conserving momentum scattering amplitudes. This procedure has confirmed the existence of a $T$-linear term in resistivity and the correlation between its strength and the size of superconductivity pairing. \cite{Shahbazi15}
  
 These  findings have motivated the present  work devoted both experimentally and theoretically to  the influence  of a finite elastic life time on the $T$-linear resistivity   in the Bechgaard salts. On the theoretical side  impurity scattering, assumed to be absent so far    in the renormalization group calculations of the scattering amplitudes, is a source of pair breaking for both unconventional superconductivity and density-wave correlations. We will show that this affects to a certain extent the constructive  interference between both types of correlations at low energy. Although this can  be sufficient  for    the complete suppression of $T_c$, it turns out that SDW correlations that build up at higher energy are less affected by disorder induced pair-breaking  and   remain relatively strong in amplitude. As the source of umklapp scattering for carriers in the  Boltzmann equation, these correlations sustain to a great extent the non Fermi liquid character of  resistivity  in spite of    the scaling  breakdown   between the $T$-linear   term   and $T_c$.

On experimental side, previous  investigations have shown that \tc in the \tms2x series  is  strongly affected by disorder of non magnetic origin\cite{Choi82,Tomic83a,Joo04,Joo05}, providing in turn some hint for the existence of a non conventional superconducting  gap displaying both signs over the Fermi surface in this  spin singlet superconductor\cite{Shinagawa2007}, pointing for instance  to  $d$-wave (or $g$-wave) type of superconductivity. As the linear term of transport at low temperature was found to be related to $T_c$,  we feel it is legitimate to test the robustness of the established connection between $A$ and \tc when \tc can be modified by other means but pressure, namely by the influence of non magnetic disorder \cite{Suzumura89}. It is worth mentioning that in
 {$\mathrm{Sr_2RuO_4}$, an equal sensitivity to non magnetic disorder has been  observed\cite{Mackenzie98} suggesting that  spin triplet pairing is a  possible candidate for the SC ground state in this material\cite{Maeno94}.

Fortunately, \tmc\, is the unique  system among other superconductors in the \tms2x series in which structural disorder  can control the elastic scattering time. This property is based on the peculiarity of the tetrahedral symmetry of the anion \cl. 

The structure of \tm2x salts belongs to the triclinic space group $P\overline{1}$ with every anion site  located on an inversion center. Hence, as long as the local symmetry of  anions like   \pf\  is centrosymmetric, their actual orientation  always fits  the overwhole crystal symmetry. Thus,   no additional disorder is  introduced. However, non centrosymmetric anions such as tetrahedral  \C or \R may have their oxygen atoms pointing towards  methyl groups or  Se atoms of one or the other  neighboring donor molecules\cite{Pouget96,Lepevelen01}. These (at least two) possible directions  lead in turn to a potential source of disorder. While there is no disorder on average at elevated temperature since these tetrahedral anions are rotating as shown by NMR data\cite{Takahashi84},  the minimization of  entropy due to the reduction of degrees of freedom  in thermodynamical equilibrium triggers an anion ordering at low temperature with the concomitant occurrence of a  superstructure. In addition,  the relative orientation at low temperature of neighboring anions depends on the nature of the anions. For (TMTSF)$_2$ReO$_4$, the orientation of \R  alternates along the three directions  providing a periodic lattice distortion of  wave vector $(1/2,1/2,1/2)$ (in units of reciprocal lattice vector in each direction) with a concomitant  metal insulator transition below \TAO=176~K \cite{Moret82}, whereas the alternation of \C occurs solely along the $b$ axis in (TMTSF)$_2$ClO$_4$  below \TAO= 24~K, leading in turn to the (0,1/2,0)  order \cite{Pouget83b,anioncavity}. Such a doubling of periodicity along $b$ is thus responsible for the doubling 
of the  Fermi surface folded along $b^\star$  in slowly cooled samples. These samples  retain in turn metallic properties at low temperature.

A possible means of introducing controllable disorder  in an otherwise well ordered \tmc crystal  is the substitution  of \cl\ by \R anions.  The consequences of the isostructural anion alloying on the electronic properties such as superconductivity have been revealed  long ago\cite{Tomic83a}, even though the actual structural reasons were not elucidated. For instance, \R replacing \C  in \tmc  affects \tc  so severely that a 1\% \R concentration  is large enough to shift \tc from 1.3~K to 0.9~K and 10\% suppresses \tc totally\cite{Tomic83a}. Additional  investigations of structural and electronic properties  of \tmxns have been conducted subsequently\cite{Ravy86,Joo04}. What structural studies have revealed in the \C alloys at low concentration of \R  is the existence of domains in which the (0,1/2,0) \C order persists, the size of them controlling the  carrier elastic mean free path\cite{Ilakovac97}.

Another way to create defects in \tmc is to beat the slow anion orientation process by rapid cooling of the sample, as first detected by NMR, specific heat, EPR and transport measurements\cite{Takahashi83,Garoche82a,Tomic82}. A recent reinvestigation of magnetic and transport properties of \tmc under a well controlled cooling procedure in the vicinity of  the anion ordering temperature has revealed at increasing  cooling speed above 1~K/mn a cross over from a rather  homogenous localized disorder to a granular situation in which anion-ordered puddles are embedded in an anion-disordered background\cite{Yonezawa18}. This picture is reminiscent of the present situation in alloys where the carrier mean free path of the well ordered grains is also limited by their own size. 

The fact that the elastic  mean free path of \tmc can be controlled by disorder 
is a remarkable property of the material and  makes \tmc a unique system where to check simultaneously  the effect of a change of the mean free path on  \tc and on  quantum criticality behavior possibly  related  to the onset of superconductivity\cite{Doiron09}. 

The present report attends to fulfill this goal with a quantitative study of both superconducting and metallic state properties in two situations where disorder can be introduced with a concomitant control of the mean free path. First, \textit{via} studying   lightly \R substituted \tmc\, samples and second the fast cooling of pure \tmc. Theoretically, we broaden the RG  approach of the quasi-1D electron gas model to include impurity pair breaking effects on the renormalization of umklapp scattering  that enters   the linearized Boltzmann equation of electrical transport.

  Our study will show that the correlation between the non-Fermi liquid resistivity and $T_c$,  which is  well established under pressure\cite{Doiron09}, breaks down when \tc in this d-wave superconductor is suppressed by non-magnetic disorder. This result emphasizes the strong-pair breaking role of a limited elastic mean free path  on \tc  as opposed to the much weaker influence on antiferromagnetic fluctuations, which are believed to be a major ingredient for pairing in this superconductor. 

 In Sec.~II, the low temperature transverse resistivity measurements of (TMTSF)$_2$(ClO$_4$)$_{1-x}$(ReO$_4$)$_x$ alloys are presented and their superconducting and non-Fermi liquid properties described. Sec.~III is devoted to the results of the numerical solution of the linearized Boltzmann equation  of resistivity for the Q1D electron gas model with umklapp scattering and pair breaking, as yielded by   the renormalization group method   detailed  in the Appendix A. In Sec.~IV,  we discuss the results and their  connection with theory. In Sec.~V, we summarize and conclude this work. 

\section{Experiment}
The present study is based on transport measurements of \tmxns single crystals   grown electrochemically by Prof. K. Bechgaard in Copenhagen. All data have been taken at Orsay except for one \tmc sample already measured in a previous study at Kyoto\cite{Yonezawa18} for its superconducting properties. 

The measurement of the transverse resistivity $\rhoc$ along the least  conducting $\cstar$  direction has been privileged  because of the requested quantitative comparison between samples with different \R concentration. The frequent cracks occurring on cooling for the resistivity along $a$ or $b$ axes prevent any comparison between samples whereas $\rhoc$ is known to be less influenced by such cracks even after multiple cooling processes\cite{Cooper86}. In addition, magnetoresistance measurements have shown that the band theory should apply below 10~K as also supported by the existence of a Drude edge along the $\cstar$ axis in the same temperature range\cite{Henderson99}. A previous comparative study of transport along $a$ and $\cstar$ has shown that in spite of a very large anisotropy,  electron scattering times measured along $a$ or $\cstar$ exhibit a  similar temperature dependence at least up to 30~K\cite{Auban11a}. 

For \tmc samples studied in Kyoto, the resistivity was measured with the resistivity option of a PPMS with a dc current of 10$\mu$A reversed to cancel thermoelectric voltages [47]. For  \tmx studied in Orsay, a lock-in amplifier technique was used with an ac current of 10~$\mu$A [30,31].

One of the major difficulties in interpreting low temperature transport in most \tms2x superconductors is the existence of a negative curvature in the $R (T)$ curve approaching \tc   below 4~K\cite{Jerome80}. This negative curvature seems to be  fairly robust as it can be  observed even when SC is suppressed either by a magnetic field or in alloyed samples\cite{Auban11}. While precursors due to SC could contribute to a possible collective paraconductive contribution up to about 1.2~\tc\cite{Auban11}, there is still much uncertainty as to the origin of paraconductivity up to 3.5-4~K remaining a pending problem. Since precursor effects will not be discussed in the present article the analysis of the $\rhoc$$ (T)$ curves has been limited to the range 4-10~K  since our purpose  is a study of the single particle non-Fermi liquid properties only.

The study of transport  has been conducted in the  alloys series \tmx where the size of the ordered regions is controlled by the amount of the \R substituent. The resistivity data are displayed on Fig.~\ref{rhocvsT.pdf}. We also emphasize that the samples on Fig.~\ref{rhocvsT.pdf} have been labelled according to the nominal concentration of \R which may differ significantly from the results obtained  in an electron microprobe analysis\cite{microprobe}.

The resistivity of \tmc is \textit{at variance} with that of  \tmp6 since no clear $T$-linear  resistivity behavior is emerging from the raw data of \tmc\cite{uniax}.

\begin{figure}[h]
\includegraphics[width=8cm]{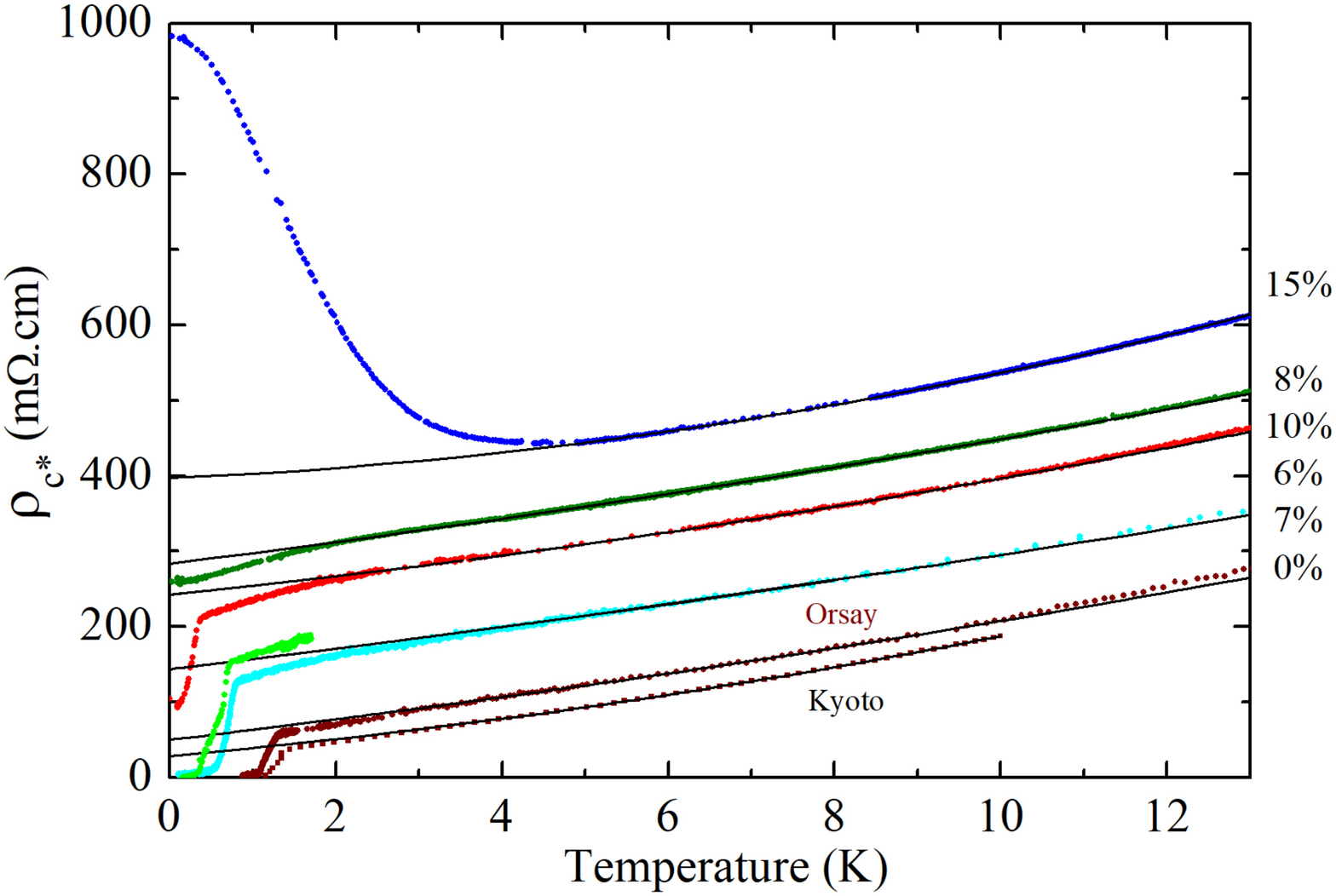}
\caption{
Interlayer resistivity  data for the \tmx solid solutions all measured in the relax state and normalized to $28\, \Omega\cdot$cm   at ambient temperature. Samples have been labelled according to their nominal \R concentration which may deviate in some cases from the actual defects concentration given by an electron microprobe analysis which in turn agrees with  the value of the  residual resistivity. The pure sample measured at Kyoto displayed a slightly smaller residual resistivity ($30\,{\rm m}\Omega\cdot$cm)  than the Orsay one, ($50\,{\rm m}\Omega\cdot$cm). This is in line with \tc= 1.35~K at Kyoto while \tc is only 1.18~K in the Orsay sample. }
\label{rhocvsT.pdf}
\end{figure}

Note that every sample on  Fig.~\ref{rhocvsT.pdf}  differs by the  behavior  of their resistivity at low temperature. Both 0\% samples exhibit a complete superconducting transition although their \tc and their residual resistivity $\rho_{c^\star0}$ are slightly different. The SC transition is still complete in the 7\% sample. We also notice a  complete  transition for the 6\% sample (green curve in Fig.~\ref{rhocvsT.pdf}) but the broadening at low temperature is a signature of a  proximity effect between superconducting puddles\cite{Yonezawa18}. A further depression of \tc is observed in the 10\% sample, but there the finite zero temperature value of the resistivity suggests the existence of superconducting islands far form each other precluding global SC coherence through proximity effect. The 8\% indicates the absence of any transition at finite temperature. The value of its $\rho_{c^\star0}$ appears to be reliable, although a tiny break in the $\rhoc(T)$ data (not visible on Fig.~\ref{rhocvsT.pdf}) occurring around  5~K made the polynomial analysis of the temperature dependence  less reliable.   In the 15\% 
sample a metal-insulator transition is observed at 2.5~K which can be related to a SDW state occurring in anion-disordered sample after fast cooling\cite{Tomic83a}. Similarly, the  sample with a nominal concentration of 17\% (not shown on Fig.~\ref{rhocvsT.pdf} displays the onset of a SDW at 3.9~K.

The  residual resistivity $\rho_{c^\star0}$  can be extracted relatively easily at different \R concentrations from the data on Fig.~\ref{rhocvsT.pdf}, as it is only weakly dependent on the determination of the other prefactors in the polynomial fit (\textit{vide infra}),  but the correct analysis of the inelastic temperature dependence is a very delicate operation even in pure samples as already discussed in several previous publications\cite{Doiron09,Auban11a}.

The present study utilizes a 
polynomial  fitting procedure  keeping $A$ and $B$  fixed  within the fitting window for each samples on Fig.~\ref{rhocvsT.pdf}. As it is difficult to put error bars on such fits we have privileged the data coming from  three different fitting ranges between 4.5~K and 10~K leading to values for $\rho_{c^\star0}$, $A$ and $B$ from the temperature dependence on Fig.~\ref{rhocvsT.pdf} which are displayed on Fig.~\ref{ATCvsrho.pdf}. The fitting window has been adequately chosen in order to avoid  the onset of the as yet uncharacterized paraconductive contribution at low temperature and the proximity of the anion ordering at high temperature\cite{paracond}.
\begin{figure}[h]
\includegraphics[width=8cm]{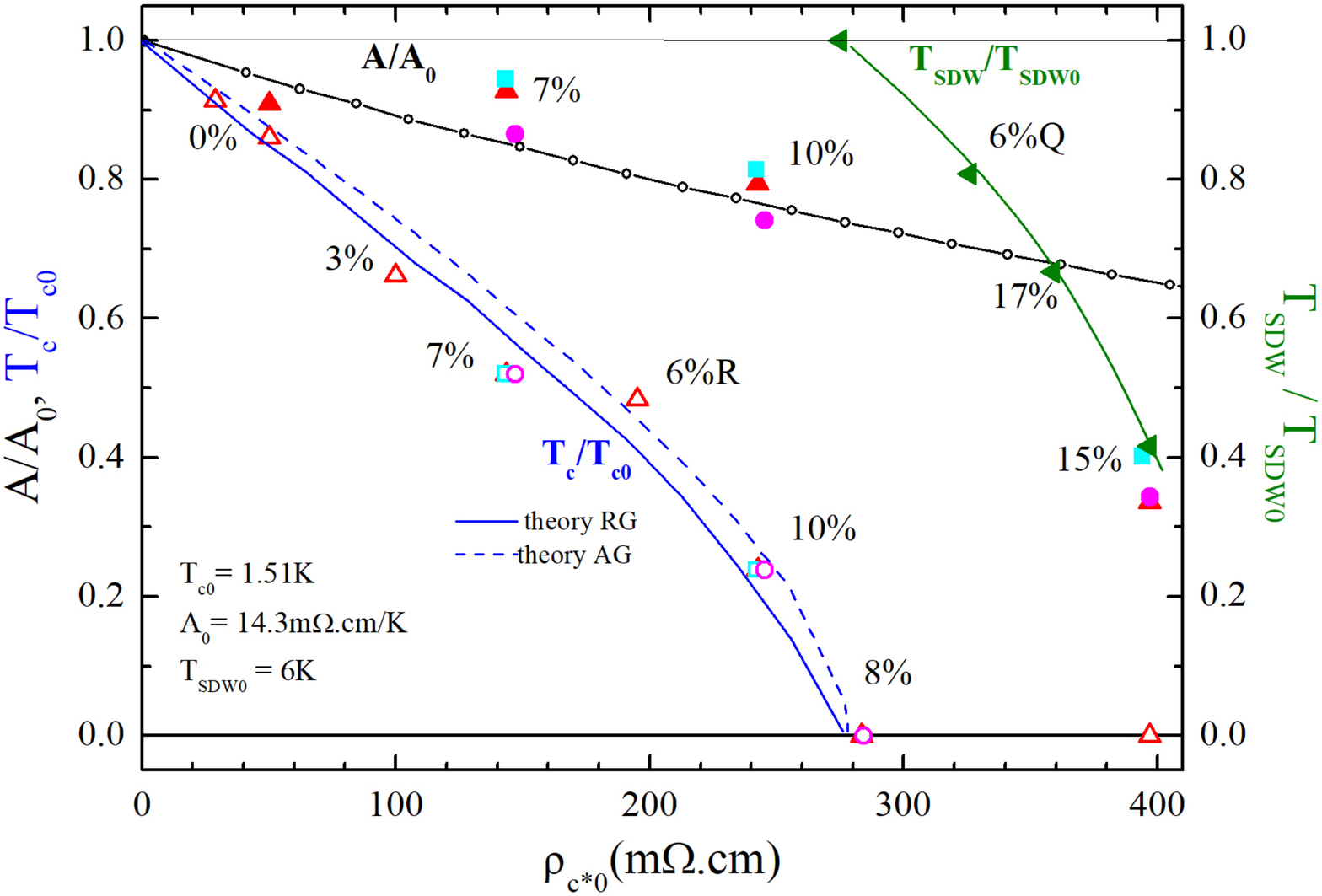}
\caption{\tc for the onset of superconductivity from experiment and theories and   \tsdw versus the residual resistivity $\rho_{c^\star0}$. Normalized value of $A$ from  the theory (open circles) and from the fits with different fitting ranges (red triangles 4.5-10~K, blue squares 5-9~K, magenta circles 5-10~K). $A$ has been normalized to the value linearly extrapolated to  $\rho_{c^\star0}$= 0.}
\label{ATCvsrho.pdf}
\end{figure}
 Transition temperatures defined by the onset temperatures have been normalized to our highest observed transition for superconductivity namely, 1.51~K and to 6~K for the SDW  transition as observed  in case of strong anion disorder\cite{Gerasimenko14}. All samples on Fig.~\ref{ATCvsrho.pdf} refer to the very slowly cooled regime except for the 6\% which has been measured both under    relax and  quenched conditions. The data of $A$ for the 15\% sample are departing from the  slow decrease with $x$  observed at lower concentrations. This may actually be the consequence of the resistivity upturn due to the SDW phase leading in turn to an underestimation of $A$ using the present fitting window.
 
 The $x$-axis of Fig.~\ref{ATCvsrho.pdf} refers to the residual resistivity as shown on the fits of Fig.~\ref{rhocvsT.pdf} and not to the actual scattering rate. We do not expect this resistivity to be directly related to  the  elastic scattering rate whenever the conducting medium departs from homogeneity. This is likely true for the present data  at high \R concentrations when the superconducting transition becomes incomplete at say, 10\%. 
 
In the solid solution  \tmxns two different anion order are competing according to x-ray diffuse scattering experiments\cite{Ravy86,Ilakovac97}. The (1/2, 1/2, 1/2) \R ordering is responsible for a metal-insulator transition at 176~K in \tmtsfreo4\cite{Moret82} while the  (0,1/2, 0) \cl ordering at 24~K folding the  Q1D Fermi surface 
 leads to a metallic phase at low temperature with a reduced value of the residual resistivity\cite{Yonezawa18}.  At concentrations  for $x$ between 0 and 1, both anion orders are competing, but on the \cl\  rich side of the solid solution (0,1/2, 0) long-range order of \cl persists up to about $x = 3\%$\cite{Ravy86,Ilakovac97}. Increasing the \R content further, the \cl\ ordering becomes short-ranged above 7\% or so and a SDW ground state is stabilized instead of superconductivity\cite{Tomic83a}. The picture which can be inferred from the x-ray analysis is thus an homogenous medium with local  
 impurities  at very low \R concentration only becoming heterogenous above 3\% where  each impurity is surrounded by a disordered volume. The mean free path  has been shown to be very large, of order of 1600 nm in  a pristine sample\cite{Yonezawa08a,Yonezawa18}. As the size of the ordered domains in alloys is likely to be much smaller than 1600 nm, we may assume that the mean free path  in alloys is determined by the typical domain size. The predominant  role of the domain size limiting the mean free path is also corroborated by a similar study performed on the solid solution \tmasf6sbf6  when the symmetry of the anions is central   which has shown that even a 15\% concentration of \sb~has no detectable effect on \tc\cite{Ola}, being \textit{at variance} with the present alloying effect in (TMTSF)$_2$ClO$_4$.
 
 The Abrikosov-Gorkov  theory\cite{Abrikosov61} of a conventional superconducting $T_c$ in the presence of pair-breaking has also been reported on Fig.~\ref{ATCvsrho.pdf} together with the results of the renormalization group theory (Sec.~III and Appendix~A). We observe a satisfactory agreement between the onset temperatures and theories when the residual resistivity is considered as a representative of the scattering rate\cite{Suzumura89}. Such an agreement, however,  may be  surprising as the residual resistivity  is the macroscopic resistivity of the entire sample which we found to become heterogenous at high \R concentration, while \tc is the onset transition temperature of  well ordered domains. Consequently, 
this agreement between experiment and theories infers in turn the existence of some connection, $\rho_{c^\star0} \propto 1$/$\lambda$, 
between  the mean  free  path $\lambda$ determined by domain size  entering the theory and the macroscopic resistivity.

\section{Theory: renormalized linearized  Boltzmann equation for resistivity}
\label{Theory}

In this section we sketch out the Boltzmann-RG approach to the calculation of electric resistivity in the presence of pair breaking. 
The calculation    combines the linearized Boltzmann equation with the renormalization group method  in the framework of the Q1D electron gas model for the Bechgaard salts\cite{Shahbazi15}. Prior to do so,  it is important to draw attention on the fact  that on experimental grounds, both longitudinal ($\Delta\rho_a$) and transverse ($\Delta \rho_{c^*}$) inelastic contribution to resistivity, though differing by a factor $\sim 10^4$ in amplitude  for a compound like (TMTSF)$_2$ClO$_4$,  exhibit essentially  the same temperature dependence  below 30~K or so\cite{Auban11a}. Such a concurrence   indicates that in  this temperature range, they are apparently governed by the same scattering mechanism. In the following we shall then focus on the calculation of longitudinal resistivity for an array of weakly coupled chains in the  $ab$ plane perpendicular to the $c$ direction.

We first consider   the Boltzmann equation of the Fermi distribution function $f_{\boldsymbol{k}}$,
\begin{equation}
\label{Boltz}
\left[{\partial f_{\boldsymbol{k}}\over \partial t}\right]_{\mathrm{coll}} = e \boldsymbol{{\cal E}}\cdot \nabla_{\hbar  \boldsymbol{k}}f_{\boldsymbol{k}},
\end{equation} 
which stands  for coherent carriers of charge $e$   coupled to an external static electric field $\boldsymbol{{\cal E}}={\cal E} \hat{a}$. The collision term on the left depends on the interparticle (umklapp) scattering and collisions to impurities or defects. The linearization of the equation standardly proceeds by looking at small dimensionless deviations $\phi_{\boldsymbol{k}}$ to the equilibrium Fermi distribution function $f^0_{\boldsymbol{k}}= 1/(e^{\beta \epsilon^p_{\boldsymbol{k}}}  + 1)$, which in leading order yields the small variation $\delta f_{\boldsymbol{k}} \simeq f^0_{\boldsymbol{k}}(1-f^0_{\boldsymbol{k}})\phi_{\boldsymbol{k}}$. Here 
\begin{equation}
\label{Spectrum}
\varepsilon^p_{\boldsymbol{k}} = \hbar v_F(pk-k_F) -2t_\perp \cos k_\perp d_\perp - 2t_\perp' \cos 2k_\perp d_\perp
\end{equation}
is the electron spectrum of the Q1D electron gas model, where $v_F$ is the Fermi velocity for right ($p=+$) and left ($p=-  $) moving carriers along the chain, $\pm k_F$ are the 1D Fermi points in the absence of interchain hopping $t_\perp$; the second nearest-neighbor interchain hopping term of amplitude $t_\perp'$ acts as an antinesting term that simulates the effect of pressure in the model and $d_\perp$ is the interchain distance along the $b$ direction.
 
 The linearization of (\ref{Boltz}) then yields 
 \begin{equation}
\label{LBoltz}
 -\left[{\partial \phi_{\boldsymbol{k}}\over \partial t}\right]_{\mathrm{coll}}= \sum_{\boldsymbol{k}'}{\cal L}_{\boldsymbol{k}\boldsymbol{k}'}\phi_{\boldsymbol{k}'}=  e \beta \boldsymbol{{\cal E}}\cdot \bm{v}_{\boldsymbol{k}}   
\end{equation}
 where $\bm{v}_{\boldsymbol{k}}$ is the carrier velocity in the $\hat{\boldsymbol{k}}$ direction. The collision operator is given by 
  \begin{align}
\label{Lop}
  \mathcal{L}_{\bm{k}\bm{k^\prime}} = &\, \mathcal{L}^u_{\bm{k}\bm{k^\prime}} +   \mathcal{L}^{\rm imp}_{\bm{k}\bm{k^\prime}}\cr
  = & \dfrac{(\pi\hbar v_F)^2}{{(L N_P)}^2} \sum\limits_{\bm{k}_2,\bm{k}_3,\bm{k}_4}  {1\over 2} \big| g_3(\bm{k}_{F}^p,\boldsymbol{k}_{F,2}^{p_2};\boldsymbol{k}_{F,3}^{-p_3},\boldsymbol{k}_{F,4}^{-p_4})\cr
   &-g_3(\bm{k}_{F}^p,\boldsymbol{k}_{F,2}^{p_2};\boldsymbol{k}_{F,4}^{-p_4},\boldsymbol{k}_{F,3}^{-p_3})\big|^2
 \frac{2\pi}{\hbar}  \delta_{\bm{k}+\bm{k}_2,\bm{k}_3+\bm{k}_4 + p\bm{G} }\cr
   &\delta(\varepsilon^p_{\bm{k}}+\varepsilon^{p_2}_{\bm{k}_2}-\varepsilon^{p_3}_{\bm{k}_3}-\varepsilon^{p_4}_{\bm{k}_4})  \cr
&\times  \dfrac{f^0_{\bm{k}_2}[1-f^0_{\bm{k}_3}][1-f^0_{\bm{k}_4}]}{[1-f^0_{\bm{k}}]}\cr & \times(\delta_{\bm{k},\bm{k^\prime}} + \delta_{\bm{k}_2,\bm{k^\prime}} - \delta_{\bm{k}_3,\bm{k^\prime}} - \delta_{\bm{k}_4,\bm{k^\prime}}) \cr
+ & {\pi\hbar v_F\over LN_P} {2\pi\over \hbar}g_{\rm imp}^2 \, \delta( \varepsilon^p_{\bm{k}} -\varepsilon^{p'}_{\bm{k}' }) (1-\delta_{\bm{k},\bm{k^\prime}}),
\end{align}
where $g^2_{\rm imp}$ is the square of the impurity scattering matrix element (normalized by $\pi \hbar v_F$) times the impurity concentration and $N_P$ is the number of transverse momentum wave vectors.  In the framework of the Q1D electron gas model, the  electron-electron  umklapp scattering amplitude is given by   $g_3(\bm{k}_{F}^p,\boldsymbol{k}_{F,2}^{p_2};\boldsymbol{k}_{F,3}^{-p_3},\boldsymbol{k}_{F,4}^{-p_4})$  (normalized by $\pi \hbar v_F$).   For an  array of quarter-filled  but weakly dimerized chains, this momentum dissipative process has its origin in the scattering of two carriers from one side of the Fermi surface to the other,  which is made possible if momentum conservation in (\ref{Lop}) involves
 the longitudinal  reciprocal lattice vector $\bm{G}= (4k_F,0)$ at half filling ($k_F=\pi/2a$). Umklapp  scattering amplitude is  evaluated on the Fermi surface sheets $\bm{k}_F^{p=\pm}=(k^p_F(k_\perp),k_\perp)$, which is determined by the equation  $\varepsilon^p_{\boldsymbol{k}_F}=0$ and parametrized by the transverse wave vector $k_\perp$. As we will see below,  its momentum dependent renormalization   at temperature $T$, will be  obtained by the RG technique at the one-loop level (see also Appendix A). 
 
 The longitudinal electric current is given by the expression in leading order
 \begin{equation}
\label{ }
j_a \simeq {2e\over L N_P d_\perp}\sum_{\bm{k}} v_F f^0_{\bm{k}}(1-f^0_{\bm{k}}) \phi_{k_\perp},
\end{equation}
where  $\phi_{\bm{k}_F}\equiv \phi_{k_\perp}$ can  been taken on the Fermi surface at low temperature. The conductivity $\sigma_a$ or the inverse of the resistivity along the chains can then be written in the form
\begin{equation}
\label{BRGrho}
\sigma_a =\rho_a^{-1} = {e^2\over c \hbar} \langle \bar{\phi}_{k_\perp} \rangle_{{\rm FS}},
\end{equation}
where we have rescaled the deviation ${\bar{\phi}_{k_\perp}= \phi_{k_\perp}/(\beta e {\cal E} d_\perp)}$ and inserted the lattice constant $c$ in the $c^*$ direction. Here $\langle ... \rangle_{\rm FS}$ corresponds to an average over the Fermi surface. Therefore the numerical solution of (\ref{LBoltz}) for  $\bar{\phi}_{k_\perp}$ using the momentum and temperature dependence of $g_3$ provided by the  RG method leads to the resistivity as a function of the temperature\cite{Shahbazi15}. 

We  can complete the standard description of interactions in the   Q1D electron gas model entering in the RG  calculations of Appendix~A.  Thus besides the spectrum (\ref{Spectrum}) and the aforementioned umklapp term, we have the backward and forward scattering amplitudes $g_1(\bm{k}_{F,1}^-,\boldsymbol{k}_{F,2}^{+};\boldsymbol{k}_{F,3}^{-},\boldsymbol{k}_{F,4}^{+})$ and  $g_2(\bm{k}_{F,1}^+,\boldsymbol{k}_{F,2}^{-};\boldsymbol{k}_{F,3}^{-},\boldsymbol{k}_{F,4}^{+})$, also  defined on the Fermi surface. 
For a superconducting compound like (TMTSF)$_2$ClO$_4$, which at ambient pressure is close to a magnetic QCP, typical  values of the bare parameters of the model have been  assessed  in previous works \cite{Sedeki12,Shahbazi15,Shahbazi16}. For example the Fermi temperature $T_F= \hbar v_F k_F/k_B\simeq 3000$~K and interchain hopping $t_\perp/k_B \simeq 200$~K are representative band parameters for the Bechgaard salts\cite{Grant83,Ducasse86}. The bare couplings (normalized by $\hbar \pi v_F$) we shall use are $g_1=g_2/2\simeq0.32$ and   $g_3\simeq 0.033$, consistently with uniform susceptibility measurements  and weak dimerization of the organic  stacks. In these conditions the RG calculations for relatively weak antinesting parameter $t_\perp'$ lead to  SDW ordering temperatures that are compatible with those found in the  antiferromagnetic Bechgaard salts at ambient pressure.  

To suppress the SDW state and bring the system  close to the QCP,  $t_\perp'/k_B$ is raised to 42~K or so, yielding a  critical temperature $T^0_c\sim 1$~K for (d-wave) superconductivity\cite{Shahbazi15}, which is congruent to the situation that prevails  in  (TMTSF)$_2$ClO$_4$ in slow cooling.   This set of figures  will fix the initial conditions with zero pair breaking effects ($\tau_0^{-1}=0$) in the  RG  calculations [Eqs.~(\ref{Flowg})]  and    the numerical solution of the Boltzmann equation (\ref{LBoltz}). Finally  a  normalized impurity scattering matrix element $g_{\rm imp}^2=0.001$ has been used in order to simulate a small residual resistivity term from (\ref{Lop}); its value smoothly increases  with $\tau_0^{-1}$. It is worth noticing that the interplay between the impurity and inelastic parts in (\ref{Lop}) has been found to be negligible.

\begin{figure}
 \includegraphics[width=8cm]{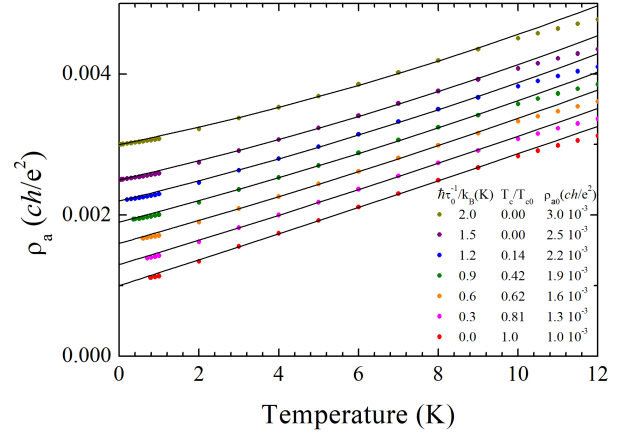}
\caption{Calculated temperature dependent resistivity as a function  of pair-breaking parameter $\hbar \tau_0^{-1}/k_B$. The continuous lines are fits to the polynomial expression $\rho_a(T) = \rho_{a0} + AT + BT^2$. }
\label{rhoa.pdf}
\end{figure}

In Figure~\ref{rhoa.pdf} we show the resistivity calculated from Eqs. (\ref{BRGrho}) and (\ref{Flowg}) in the low temperature range down to  the onset of critical superconducting domain occurring  close  to $T_c$  where the renormalization flow goes to strong coupling. 

Thus for zero pair breaking ($\tau^{-1}_0=0$), the system is very close to the QCP and the resistivity becomes essentially  $T$-linear up to 10~K or so where a crossover to a sublinear temperature dependence takes place (See \cite{Tsquare}).  
The origin $T$-linearity  stems from the anomalous temperature dependence of  umklapp scattering  along the Fermi surface. Actually, despite the absence of SDW long-range order, spin correlations remain important and keep growing as  temperature is lowered\cite{Bourbon09}. This derives from Cooper pairing terms in the RG flows  (\ref{Flowg}) which interfere constructively with the electron-hole (Peierls) pairing ones  and then SDW fluctuations. This  quantum fluctuation effect results in an anomalous  temperature dependence of  umklapp scattering that transforms  the Fermi liquid quadratic $T^2$ dependence of resistivity into a $T$-linear behavior. Roughly speaking, the  resistivity goes like $\rho_a \sim T^2 \langle g_3^2 \rangle_{\rm FS}$  at  low temperature where the mean-square value of umklapp scattering on the Fermi surface, which is representative  of spin fluctuations,  goes as $1/T$ at the critical $t_\perp'$. 

\begin{figure}[h]
 \includegraphics[width=8cm]{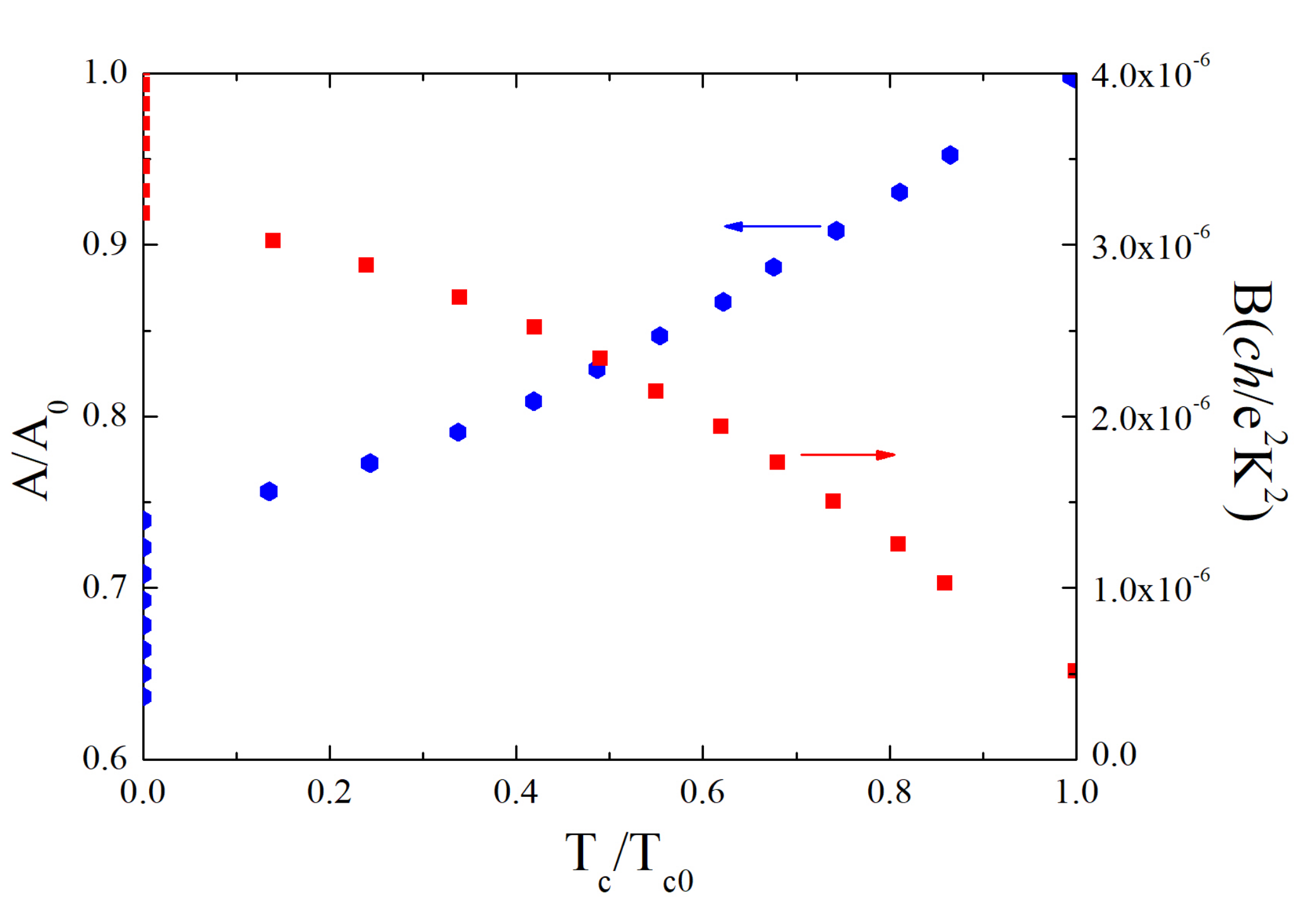}
\caption{$A$ (left scale) and $B$ (right scale) coefficients of the polynomial fit $\rho_a(T) = \rho_{a0} + AT + BT^2$ of the calculated resistivity (Fig.~\ref{rhoa.pdf}) as a function of $T_c$. }
\label{AB}
\end{figure}

When pair-breaking $\tau_0^{-1}$ departs from zero, $T_c$ gradually decreases and follows relatively closely the Abrikosov-Gorkov mean-field result down to the critical $\hbar\tau_0^{-1}/k_B\simeq 1.25$~K where $T_c\to 0$ (see Fig.~\ref{ATCvsrho.pdf}). Concomitant with the fall of $T_c$, the temperature dependence of resistivity  shows comparatively slow, albeit noticeable, alterations. We have proceeded to the fit of calculated resistivity curves of Fig.~\ref{rhoa.pdf} to the polynomial form, $\rho_a =\rho_0 + AT + BT^2$, in the interval from 10~K down to the lowest temperature. According to Figure~\ref{AB}, the linear coefficient $A$ of resistivity weakens with the size of $\tau_0^{-1}$ or the reduction of $T_c$. When $T_c\to 0$   as we approach the threshold $\tau_0^{*-1}$, the reduction in the ratio $A/A_0$ reaches about 25\%; it continues a steady decrease beyond that point. As for the Fermi liquid $BT^2$  term of resistivity,   it becomes visible and weakly grows,  but regularly for $T_c/T_c^0<1$.

It follows that when pair breaking is present the scaling of $A$ with $T_c$ no longer holds. It must be stressed, however, that the threshold energy scale for pair-breaking $\hbar \tau_0^{*-1}$, which is of the order of $ k_BT_c^0$,  only affects electronic states within a  narrow energy shell around the Fermi surface. Although the corresponding decline of the Cooper loop (${\cal L}_C$) entering the RG flow (\ref{Flowg}) is sufficient to suppress the instability against $d$-wave superconductivity,  scattering processes  taking place at higher energy distance from the Fermi surface remain virtually unaffected. These  still contribute to the mixing of Cooper pairing and  SDW correlations. The  latter alongside umklapp scattering continue to be strong and  are responsable for sustaining the amplitude of the $T$-linear coefficient of resistivity despite  the absence of a finite $T_c$.

\section{Discussion}
\label{Discussion}
Before we proceed with a comparison between theory and experiments in these alloys, it is important to notice that the present study is not dealing with a random distribution of strongly localized defects. Instead, we are facing a situation where the carrier life time in ordered  regions is governed by the size of extended  of disordered domains.  The transport data showing the remnence of non superconducting domains at high \R concentration suggest we are facing a problem of granular materials. Hence, the value of the residual resistivity cannot  be simply extracted from the knowledge of the \R concentration. The situation in the solid solution  has similarities with that encountered in  pure \tmc when the disorder is caused by rapid cooling of the sample\cite{Yonezawa18}. Therefore, we shall use a similar procedure to extract  the  fraction of the respective ordered and disordered regions from the resistivity data according to the effective-medium theory for a two-component conductor\cite{Landau,Creyssels2017},
\begin{eqnarray}
{1/\rho_{\mathrm{eff}}=\left( p(1/\rhoMin)^{1/3}+(1-p)(1/\rhoMax)^{1/3}\right)^3},
\label{rhoeff}
\end{eqnarray}
where $p$ is the volume fraction of the anion ordered domains. We take $\rhoMin= 50~{\rm m}\Ohmcm$ according to the data on Fig.~\ref{rhocvsT.pdf} and $\rhoMax$ is given by $\rhoMax = \rhoMin + \Delta\rhoc = 280~{\rm m}\Ohmcm$ where $\Delta\rhoc$ is the drop in residual resistivity coming from the anion ordering at 24~K which is known from the pure \tmc \,data ($\Delta\rhoc$= $230~{\rm m}\Ohmcm $)\cite{Yonezawa18}. The derived values for $p$ are displayed on Fig.~\ref{pvsrhoc}. Using the value for the residual resistivity of sample 0\% in Eq.~(\ref{rhoeff}) would provide  $p$=1 for that sample in which the residual defects are not under control. As x-ray data have shown that full anion ordering does not exist even in pure and the best relaxed samples\cite{Pouget12,rhopf6}, we feel entitled to proceed differently for the determination of $p$  in sample 0\%. 
The derivation of $p$ from Eq.~(\ref{rhoeff}) is expected  to become meaningful only in samples where the additional disorder provided by alloying is  large compared to the residual disorder in a pristine sample. Therefore we have proceeded for  $p$ of the  0\% sample with a linear interpolation between $\rho_{c^\star0}$= 0 and  $100~{\rm m}\Ohmcm $ ($x=3\%$) on Fig.~\ref{pvsrhoc} leading to the determination of $p$ ($\simeq 0.78$) in the sample 0\% of residual resistivity $\rho_{c^\star0}$= $50~{\rm m}\Ohmcm $.

From Fig.~\ref{rhocvsT.pdf}, the 10\% sample   exhibits a superconducting transition ending at a finite value for the resistivity at zero temperature, which means that ordered regions are not percolant. Furthermore,  in the 6\% corresponding to $p =0.18$ according to Fig.~\ref{pvsrhoc},   the transition becomes complete meaning that pairing coherence becomes infinite. The volume fraction for ordered domains in 6\% is  smaller than the actual value expected for the percolation threshold,  namely, $p =0.3$ at which the first cluster of 3D ordered anions with infinite length develops\cite{Stauffer94}.  Such a situation can be understood taking into consideration the proximity effect which enables superconducting coherence over the entire volume even though  ordered regions are still separated by small  gaps of anion-disordered matter.

\begin{figure}[h]
\includegraphics[width=8cm]{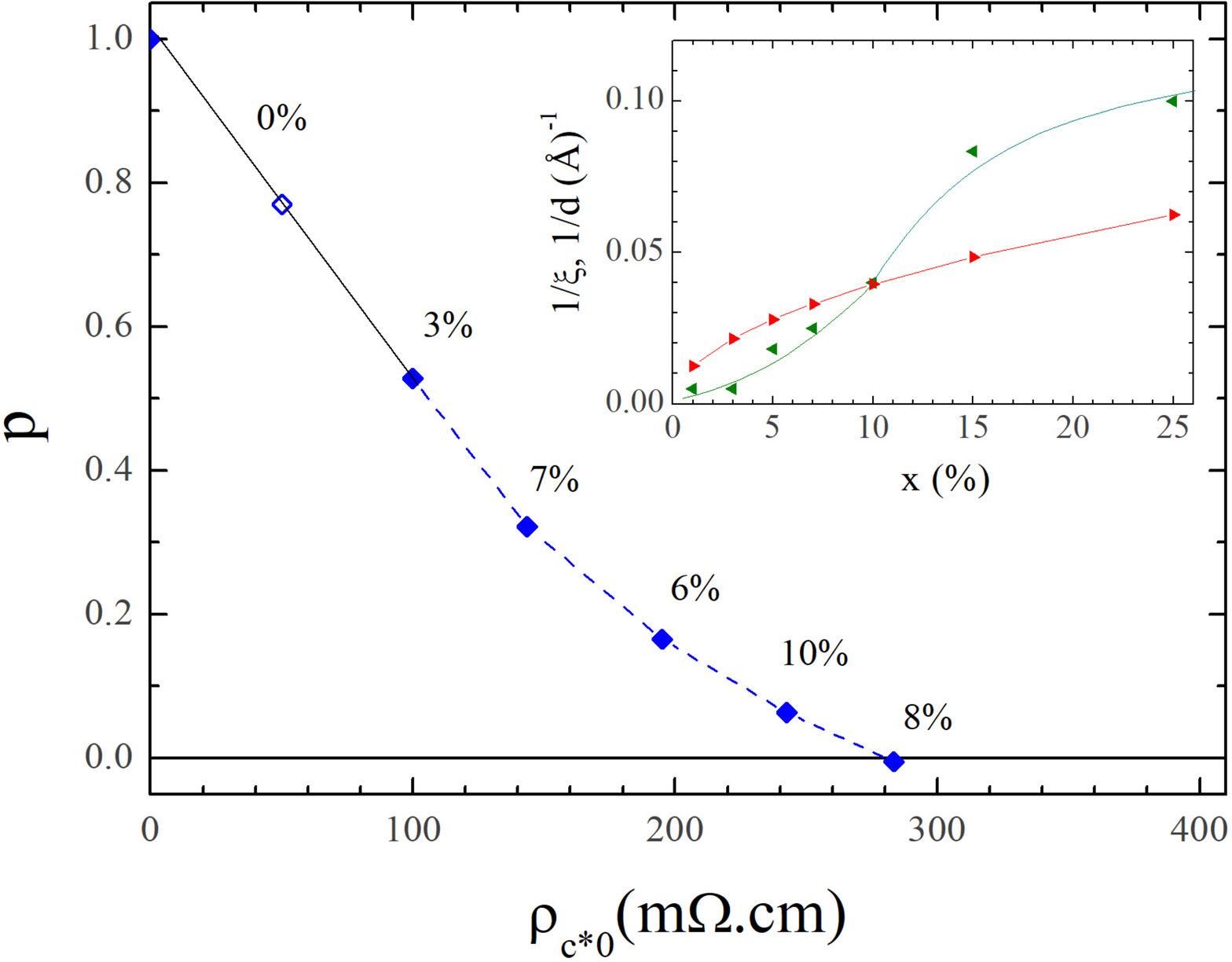}
\caption{ The anion-ordered volume fraction $p$ derived from the residual resistivity and the effective medium model. The insert shows the $x$ dependence of the inverse correlation length for the (0, 1/2, 0) ordering\cite{Ravy86,Ilakovac97} and the inverse length between \R sites located at random. We note that around $x= 10~\%$ the (0,1/2,0) anion-ordered coherence length becomes shorter than the average distance between centers. It is  above such  a concentration that a granular behavior prevails. The superconducting coherence spreads over the  whole volume in the 6~\% sample with $p = 0.18$ whereas the percolation threshold for ordered regions should arise around $p$= 0.3, \textit{i.e.}, close to the 7~\% sample. This is a consequence of the proximity effect for superconductivity.  }
\label{pvsrhoc}
\end{figure}
\begin{figure}[h] 
\includegraphics[width=8cm]{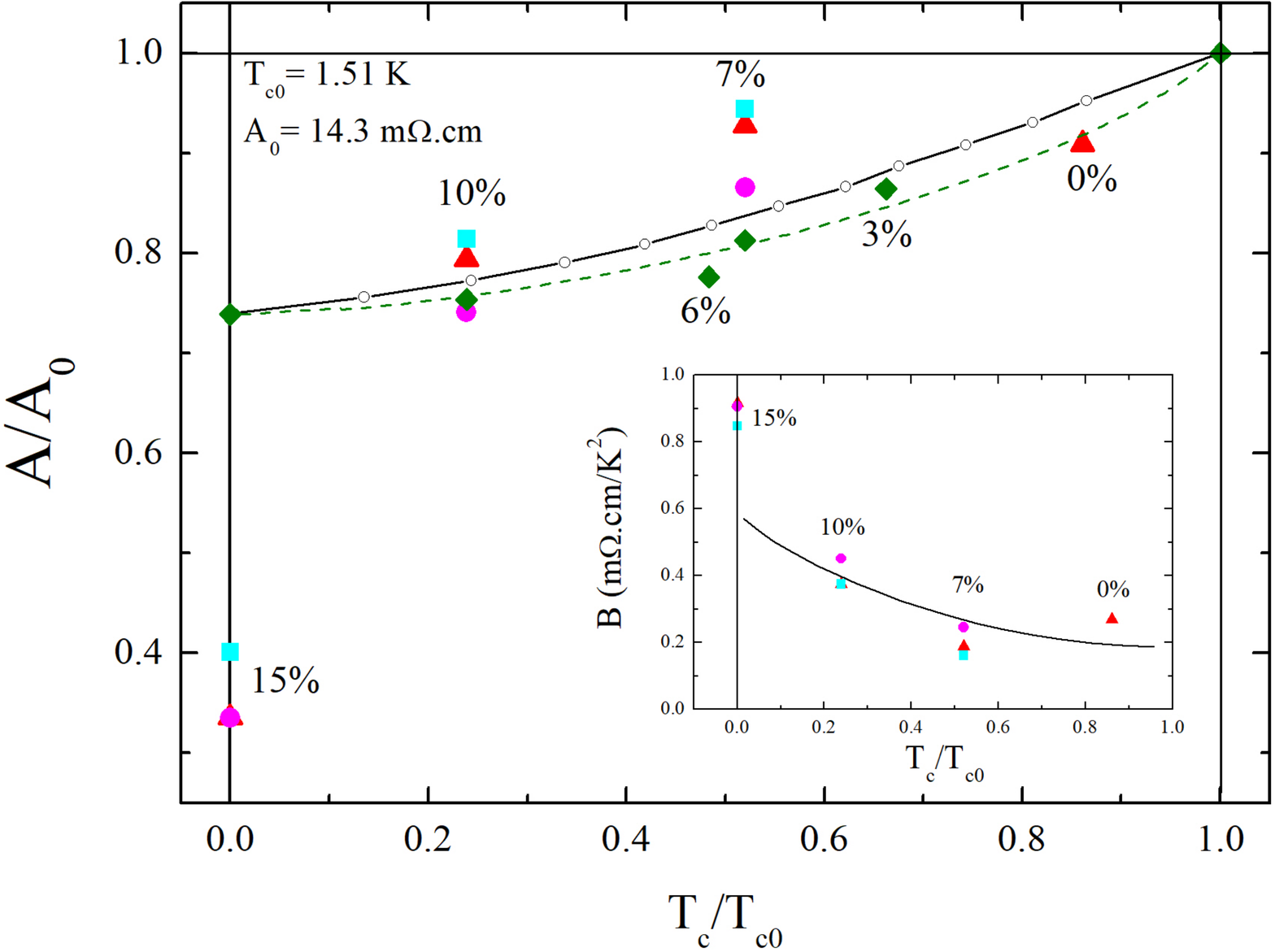}
\caption{Dependence of $A/A_0$   on $T_c$, from experiment and theory (full line). The colored square, triangle and circle symbols correspond to experimental data according to  the different  fitting ranges described in Fig.~\ref{ATCvsrho.pdf}. The green diamonds are  estimated $A/A_0$ derived from the two-conductor model  Eq.~(\ref{rhoeff}) for the inelastic contribution. $A/A_0  = 0.75$ at $T_c =0$~K and    the volume fractions $p$ according to Fig.~\ref{pvsrhoc} have been used. The green dotted  line  is a guide for the eye.  Inset, displays the $B$ dependence on $T_c$, the line is a guide for the eye.}
\label{AvsTc.pdf}
\end{figure}

Regarding the SDW transition on Fig.~\ref{ATCvsrho.pdf}, we note that the highest transition temperature at 6~K arises in a clean but already fully anion disordered sample\cite{Gerasimenko14}. Hence, in a quenched state additional scattering coming from  \R  anions is expected. This is the situation encountered  in the quenched-6~\% samples where no (0,1/2,0) order prevails because of the quenched state and   additional scattering  coming from \R substitution. As emphasized in Sec.~III,  non magnetic defects  provide alongside  antinesting some additional pair-breaking reducing density-wave correlations and then  $T_{\rm SDW}$  $-$ in  the same way as non magnetic defects act in   $d$-wave superconductors \cite{Chang86,Dora03,Ola2}.

We can now plot the dependence of $A$ on \tc, as shown on Fig.~\ref{AvsTc.pdf} comparing experimental data provided by the analysis of the resistivity of the metallic phase above \tc to the  Boltzmann-RG calculation. We must be cautious because  $A$ is related to an homogenous medium  in the two extreme cases of high and low \tc only where the sample is anion-ordered or disordered respectively but in-between (say, 7~\%) we face a more complex situation where $A$ is related to the mixture of two metallic phases whereas \tc is given by the onset of superconductivity in the ordered phase only.

 Experiments and theory of Fig.~\ref{AB} agree on the fact that the suppression of $A$ is no more than 25\% once superconductivity in alloys  is fully suppressed.  In these conditions a sustained  strength of a $T$-linear component for resistivity  indicates  that in the presence of non-magnetic pair-breaking the system moves away only slowly  from the magnetic QCP.   The amplitude of SDW correlations and with it the  enhancement of   umklapp scattering   persist to a great extent in the anomalous temperature dependence of resistivity.    Also satisfactory is the comparison between theory and experiments for the growth of the Fermi liquid  $B$ coefficient of resistivity,  as displayed  in Fig.~\ref{AB} and the inset of   Fig.~\ref{AvsTc.pdf}. The rise  of  $B$  as $T_c$ decreases is steady, but its amplitude remains small.

Similarly, in case of pure (TMTSF)$_2$ClO$_4$, the evolution of the $T$-linear term $A$ as a function of the cooling rate can be extracted from a polynomial fit of the raw data of previous works in Ref.~\cite{Yonezawa18},  which we  reproduce  in Fig.~\ref{rawdata_rho.pdf} for representative cooling rates.  The fit is in the window  between 4.5 and 10~K, namely outside the paraconductive downturn of resistivity whose amplitude increases with the cooling rate. The suppression of $A$ is at most 25\%  when the volume fraction of anion-ordered regions is monitored by the cooling rate leading to a complete suppression of $T_c$,  as displayed on  Fig.~\ref{fig6.pdf}.  

\begin{figure}[h]
\includegraphics[width=8cm]{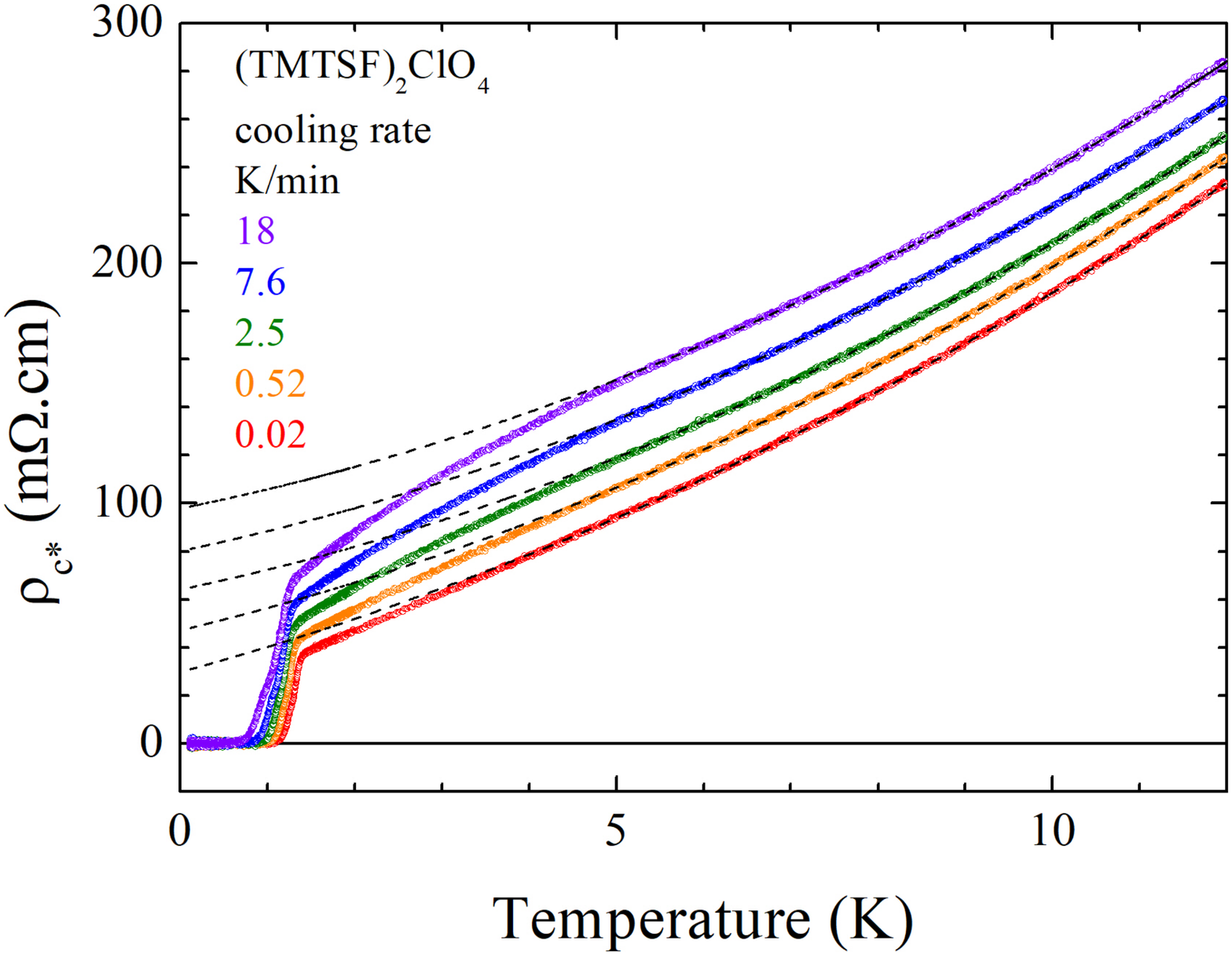}
\caption{$\rhoc(T)$ measured on \tmc  and displayed in the temperature range between 12~K and the superconducting region for 0.020, 0.52, 2.5, 7.6, and 18~K/min cooling rates across the anion ordering transition.  
The same polynomial fitting procedure as for the data on Fig.\ref{rhocvsT.pdf}  has been used to extract the $A, B$ and $\rho_{c^\star0}$ parameters at each cooling rates within the fitting window 5-10~K. The results are plotted as broken lines. } 
\label{rawdata_rho.pdf}
\end{figure}

Finally, we comment on the behaviour of   $\rhoc(T)$ in pristine \tmc  at various cooling rates  on Fig.~\ref{rawdata_rho.pdf}. These data come from the extension to finite temperature of a previous work investigating  residual resistivity and superconductivity in \tmc as the cooling rate is varied following a procedure already presented in  Ref\cite{Yonezawa18}. The resistivity has been normalized to the value of $28\, \Omega\cdot{\rm cm}$ at ambient temperature following reports of previous $\rhoc$$(T)$ measurements\cite{Joo05,Yonezawa18}. What is clear at first sight on Fig.~\ref{rawdata_rho.pdf} is the marked contrast between  the strong effect of cooling rate on  the residual resistivity and the much weaker effect on the inelastic part, a feature captured by the calculations of Sec.~III. 

The pure  \tmc\, study\cite{Yonezawa18} has shown that the material acquires a granular texture above a critical cooling rate of 1~K/mn where disordered regions separate   anion-ordered domains in which bulk superconductivity arises and possibly spread over the whole sample via proximity effect. The volume fraction $p$ related to ordered domains has been estimated at each cooling rates using the model of a two-component conductor\cite{Yonezawa18}.  The data for the fitted values of $A$ are displayed on Fig.~\ref{fig6.pdf} \textit{versus} the anion-disordered volume fraction $1-p$. $A$ has been  normalized to its value at the smallest cooling rate. We also show on the same Figure, the prediction of the effective medium model using Mathiessen's law  applied to the inelastic contribution, taking $A/A_0$= 1 and 0.75 at $1-p$ = 0 and 1, respectively, following the results from the alloys series. The full squares with the dotted line are theoretical Boltzmann-RG points corrected for the evolution of the mean free path within each puddles at concentration $p$. These values provide an improvement  for the   agreement with the  experimental data. The variation of the $T$-linear resistivity of the metallic phase above \tc at various cooling rates appears to be controlled nearly equally by  life time  and phase mixture effects.
\begin{figure}[h]
\includegraphics[width=8cm]{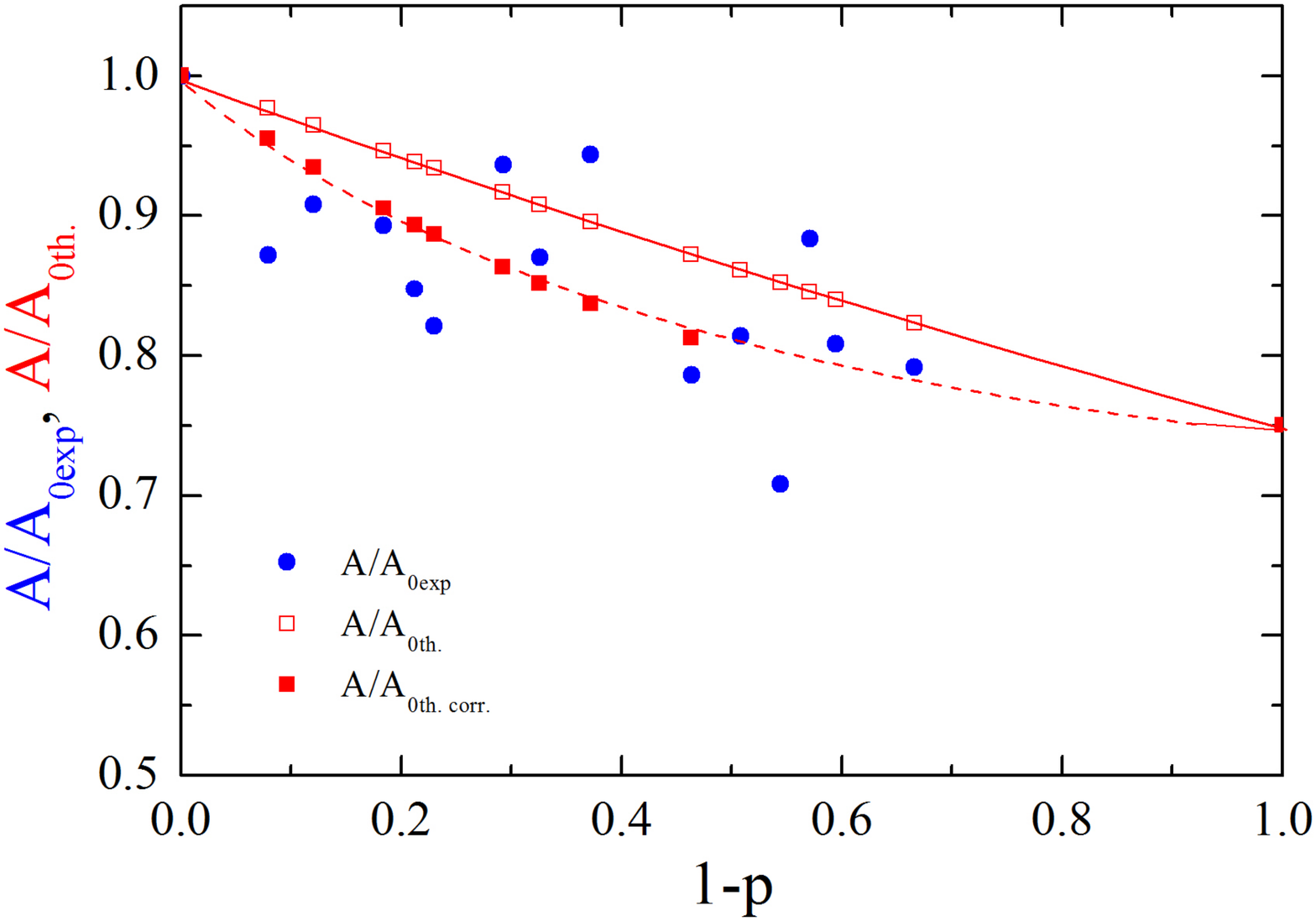}
\caption{Dependence of $A/A_0$ on   the disordered volume fraction $ 1-p$ in pure \tmc at various cooling rates, (blue dots). The model calculations  for an effective conductor with  $A/A_0=0.75$ at $p =0$ are shown as   red open squares. The model calculations  taking into account the evolution of the mean free path in anion-ordered regions are the red full squares. }
\label{fig6.pdf}
\end{figure}

\section{Conclusion}
In this work, we have examined  the influence of nonmagnetic disorder generated  by  finite domains of anion ordering on the temperature dependence of resistivity in both  (TMTSF)$_2$(ClO$_4$)$_{1-x}$(ReO$_4$)$_x$  alloys and pristine (TMTSF)$_2$ClO$_4$ at different cooling rates. These domains, presumably of different electronic structure, were  found to have a strong impact on the onset of superconducting order. $T_c$ is suppressed pointing to the existence of pair-breaking effects    congruent with unconventional Cooper pairing.  At variance with $T_c$, the pronounced $T$-linear resistivity in (TMTSF)$_2$ClO$_4$ observed  in slow cooling conditions   does not reveal as a strong decline in alloying. A polynomial decomposition of  resistivity temperature dependence  has revealed  that the linear term is not strongly weakened   with $x$, indicating that quantum critical features remain strong  despite  the suppression of superconducting long-range order. The scaling between the $T$-linear term strength  and $T_c$, which is seen under pressure in pure samples, was found to no longer hold  as a function of disorder tuned by $x$ or to a certain extent the cooling rate.

These results have been compared  to a theoretical calculation  based on the combination of the linearized Boltzmann equation of transport and the renormalization group approach to the umklapp vertex function. In the framework of the q1D electron gas model, we   have shown that a range of relatively small energy scales for pair breaking is sufficient to instill  a characteristic  fall in the metallic instability     against d-wave superconductivity.   The calculated  drop in $T_c$ 
agrees with  the one observed with $x$. Although small pair breaking energy is enough to suppress a  scale  like $T_c$, its impact on the strength of SDW fluctuations that feeds the amplitude of interparticle umklapp scattering remains relatively weak.  The anomalous growth of umklapp at low temperature was  found to be  weakly altered  along with linear resistivity   term whose amplitude remains sizable, in agreement with experiment.   Although the correlation between $T_c$ and linear resistivity apparently breaks down, the correlation of the latter with spin fluctuations, which are the core  of $d$-wave pairing, prevails.

\begin{acknowledgements}
The samples studied for this work had been synthesized and grown by the late Professor Klaus Bechgaard. This work in France  has been supported by CNRS and  in Japan by Japan Society for the Promotion of Science (JSPS) Grant-in-Aids KAKENHI Grants No. JP26287078 and No. JP17H04848. C. B. thanks the National Science and Engineering Research Council  of Canada (NSERC) and the R\'eseau Qu\'eb\'ecois des Mat\'eriaux de Pointe (RQMP) for financial support. Computational resources were provided
by the R\'eseau qu\'eb\'ecois de calcul de haute performance
(RQCHP) and Compute Canada. A. Sedeki thanks the Institut Quantique (IQ) of the Universit\'e de Sherbrooke for its financial support. 
\end{acknowledgements}

\appendix
\section{Renormalization group equations for the electron gas with  pair breaking effect}

In this Appendix  we give the one-loop RG flow equations for the momentum scattering amplitudes $g_{i=1,2,3}$ in the presence of pair breaking effects due to a finite lifetime  $\tau_0$ of the carriers.   The RG approach  to the quasi-1D electron gas is detailed in Refs\cite{Nickel06,Bourbon09,Sedeki12}.  
 Each energy shell  of thickness $\frac{1}{2}E_0(\ell)d\ell$,  located at $\pm \frac{1}{2}E_0(\ell)$ from either sides of the Fermi  sheets, is segmented into $N_p$ patches, each centered at a particular value of the transverse momentum $k_\perp$, where   $E_0(\ell)=E_0e^{-\ell}$ is the scaled bandwidth at  step $\ell$ and $E_0\equiv 2E_F$ is the initial bandwidth. The successive partial trace integration of electron degrees of freedom in the partition function leads to the renormalization or flow of the scattering amplitudes $g_i$ as a function of $\ell$. At the one-loop level, the flow combines  corrections from the electron-hole (Peierls) and electron-electron (Cooper) interfering channels of scattering. Impurity (back) scattering introduces introduce pair breaking in the form of a carrier lifetime $\tau_0$ in both $2\bm{k}_F$ electron-hole (Peierls)\cite{Patton74} and (d-wave) Cooper\cite{Sun95} channels.
This  leads to the flow equations for the $g_i(\bm{k}_{F1}^{p_1},\bm{k}_{F2}^{p_2};\bm{k}_{F3}^{p_3},\bm{k}_{F4}^{p_4}) \to g_{i}(k_{\perp1},k_{\perp 2},k_{\perp 3},k_{\perp 4})$ on the Fermi surface,  which  can be  written in the compact form
\begin{widetext}
\begin{eqnarray}\label{Flowg}
&\partial_\ell g_{1}(k_{\perp1},k_{\perp 2},k_{\perp 3},k_{\perp 4})& = (-2 g_1 \circ g_1 + g_1 \circ g_2 + g_2 \circ g_1) \partial_\ell {\cal L}_P -(g_1 \circ g_2 + g_2 \circ g_1)\partial_\ell {\cal L}_C, \nonumber \\
&\partial_\ell g_{2}(k_{\perp1},k_{\perp 2},k_{\perp 3},k_{\perp 4})& = -(g_1 \circ g_1 +g_2 \circ g_2) \partial_\ell {\cal L}_C +(g_2 \circ g_2 + g_3 \circ g_3) \partial_\ell {\cal L}_P, \nonumber \\
&\partial_\ell g_{3}(k_{\perp1},k_{\perp 2},k_{\perp 3},k_{\perp 4})& = (- g_1 \circ g_3 - g_3 \circ g_1 + g_2 \circ g_3 + g_3 \circ g_2) \partial_\ell {\cal L}_P +2 g_2 \bullet g_3 \partial_\ell {\cal L}_P,
\end{eqnarray}
\end{widetext}
where $\partial_\ell= \partial/\partial \ell$. 
${\cal L}_{\nu=P,C}$ are the  Peierls and Cooper loops whose derivative at finite temperature comprises an integration over the patch. These   take the form
\begin{align}
\label{LPC}
 \partial_\ell {\cal L}_{\nu} & (k_\perp,q_{ \perp\nu}^{(\prime)})=  \frac{E_0(\ell)}{4} {\rm Re} \Big\{\sum \limits_{\mu=\pm 1} \int_{k_\perp - \frac{\pi}{N_P}}^{k_\perp + \frac{\pi}{N_P}} {dk_\perp\over 2\pi} \cr
& \times \dfrac{\theta(\vert E_0(\ell)/2 + \mu A_\nu\vert - E_0(\ell)/2)}{E_0(\ell)+\mu A_\nu + i\hbar \tau_0^{-1}} \cr
&  \times   \Big[ \tanh[\beta E_0(\ell)/4] + \tanh[\beta( E_0(\ell)/4 + \mu A_\nu /2)]\Big]\Big\},\cr 
\end{align}
where
\begin{align}
A_\nu(k_\perp,q_{\perp\nu}^{(\prime)})=&- \epsilon_\perp(k_\perp) - \eta_\nu \epsilon_\perp(\eta_\nu k_\perp+q_{\perp\nu}^{(\prime)})\cr
& + \eta_\nu \epsilon_\perp(\eta_\nu k_{\perp2(4)}+q_{ \perp\nu}^{(\prime)}) + \epsilon_\perp(k_{\perp2(4)})\cr
\end{align}
with
$$
   \epsilon_\perp(k_\perp)= -2t_\perp \cos k_\perp d_\perp - 2t_\perp' \cos 2k_\perp d_\perp
 $$
 and  $q_{\perp P}^{(\prime)}= k_{\perp3}-k_{\perp2}=k_{\perp1}-k_{\perp4} (k_{\perp3}-k_{\perp1}=k_{\perp2}-k_{\perp4})$,$q_{\perp C}= k_{\perp1}+k_{\perp2}=k_{\perp3}+k_{\perp4} $; $\eta_P = 1$ and $\eta_C = -1$. $\theta(x)$ is the Heaviside function $[\theta(0) \equiv {1\over 2}]$.

The momentum dependence of couplings  in the discrete convolution  products  `$\circ$' over the internal $k_\perp$ loop variable on the right-hand side of (\ref{Flowg}) are in order $g(k_\perp,k_{\perp4},k_{\perp1}, k_\perp-q_{\perp P})g(k_\perp,k_{\perp2},k_{\perp3},k_\perp-q_{\perp P})$ for   the Peierls channel, $g(k_{\perp1},k_{\perp2},k_\perp,q_{\perp C}-k_\perp)g(k_{\perp3},k_{\perp4},k_\perp,q_{\perp C}-k_\perp)$ for the Cooper channel, and  $g(k_\perp,k_{\perp4},k_{\perp2},k_\perp-q_{\perp P}')g(k_{\perp1},k_\perp,k_{\perp3},k_{\perp1}-q_{\perp P}')$ for the `$\bullet$' product of the off-diagonal Peierls channel.

The integration of  Eqs.~(\ref{Flowg}) up to $\ell\to \infty$ gives the values of the moment dependent scattering amplitudes $g_i$ at temperature $T$. A singularity in the scattering amplitudes signals an instability of the electron gas against the formation of an ordering state at a critical temperature. For the band parameters and repulsive interactions  fixed in Sec.~\ref{Theory}, the most probable instabilities are against either SDW or d-wave SC are  orderings. At  low antinesting $t_\perp'$ and zero pair breaking $\tau^{-1}_0=0$, an SDW instability occurs with a singularity in both $g_2$ and $g_3$ amplitudes  at the modulation wave vector $\bm{q}_0= (2k_F,q_{\perp P}=\pi)$ and $T_{\rm SDW} \sim 10-20$K. By tuning the antinesting at the QCP near ${t_{\perp} '^*}\simeq 42$~K , the SDW  is suppressed and the instability    occurs in the d-wave superconducting channel   at $T_c^0\sim 1$~K, which is associated to  a singularity for both $g_2$ and $g_1$ couplings at  zero Cooper pair momentum $\bm{q}_0= (0,q_{\perp C}=0)$, but with a cosine modulation in $k_\perp$ space. $T_c^0$  stands as the maximum $T_c^0$ at $\tau_0^{-1}=0$ which we liken to the situation of the pure and slowly cooled (TMTSF)$_2$ClO$_4$ compound. 

By increasing $\tau_0^{-1}$ it is the weakening of the Cooper loop (${\cal L}_C$) singularity  in (\ref{LPC}) that affects the instability against d-wave superconductivity. $T_c/T_c^0$ goes down and follows relatively closely the mean-field  Abrikosov-Gorkov result\cite{Sun95} as shown in Fig.~\ref{ATCvsrho.pdf}. The concomitant impact of $\tau_0^{-1}$ on the momentum and  temperature profile of  Umklapp scattering and in turn on   resistivity as obtained from the Boltzmann equation (\ref{Lop})  is discussed in Sections~\ref{Theory} and \ref{Discussion}.

\bibliography{QCP_ClO4.bib}

\begin{thebibliography}{10}
\bibitem{anioncavity}
According to structural studies, the anion ordering inside the methyl group
  cavity is also accompanied by a shift of the anion itself either towards
  methyl groups for \tmc\cite{Lepevelen01,Alemany14} or towards Se atoms of a
  donor for \tmtsfreo4\cite{Moret82}.
\bibitem{uniax}
In contrast to \tmp6 a hydrostatic pressure does not allow to cross the QCP
  region in \tmc. The QCP in \tmc is expected to be located at a negative
  pressure. This situation has been reproduced applying a uniaxial elongation
  to \tmc along its $b'$-axis leading in turn to a decrease of the unnesting
  parameter $t_\perp'$ with the suppression of superconductivity and the
  concomitant recovery of an insulating state below 6K\cite{Kowada07}.
\bibitem{paracond}
In a previous study the paraconductive conribution has been attributed to a
  fluctuating SDW\cite{Auban11}. This is one among other possibilities which
  requires confirmation.
\bibitem{Tsquare}
The crossover to the sublinear temperature dependence contrasts with
  observation of Figs.~\ref{rhocvsT.pdf} and \ref{rawdata_rho.pdf} where the
  resistivity is rather found to evolve towards a $T^2$ behavior outside the
  quantum critical domain at high temperature.
\bibitem{Ola2}
The influence of the anion potential has been studied on the SDW state of
  \tmasf6sbf6 where the question of anion symmetry is irrelevant showing a 7\%
  decrease of the transition temperature in a 50\% solid solution\cite{Ola}.
\end{thebibliography}

 \bibliographystyle{unsrt}

\end{document}